\documentclass{article}

\usepackage{PRIMEarxiv}

\usepackage[utf8]{inputenc} 
\usepackage[T1]{fontenc}    
\usepackage{hyperref}       
\usepackage{url}            
\usepackage{booktabs}       
\usepackage{amsfonts}       
\usepackage{nicefrac}       
\usepackage{microtype}      
\usepackage{lipsum}
\usepackage{fancyhdr}       
\usepackage{graphicx}       
\graphicspath{{media/}}     

\title{Pearson's Correlation under the scope: Assessment of the efficiency of Pearson's correlation to select predictor variables for linear models

}

\author{
  Mustafa Attallah \\
  Saint Louis University \\
  Saint Louis, Missouri, USA\\
  \texttt\ mustafa.attallah@slu.edu\ \\
}

\begin{document}
\maketitle

\begin{abstract}
This article examines the limitations of Pearson's correlation in selecting predictor variables for linear models. Using mtcars and iris datasets from R, this paper demonstrates the limitation of this correlation measure when selecting a proper independent variable to model miles per gallon (mpg) from mtcars data and the petal length from the iris data. This paper exhibits the findings by reporting Pearson's correlation values for two potential predictor variables for each response variable, then builds a linear model to predict the response variable using each predictor variable. The error metrics for each model are then reported to evaluate how reliable Pearson's correlation is in selecting the best predictor variable. The results show that Pearson's correlation can be deceiving if used to select the predictor variable to build a linear model for a dependent variable.
\end{abstract}


\section{Introduction}
Linear models \cite{linearM2021} are commonly used in various fields such as medicine \cite{medicineEx}, engineering \cite{engineeringEX}, agriculture \cite{agricultureEx}, and many other significant applications. A linear model describes the relationship between a dependent variable and one or more independent variables using a linear equation. In linear modeling, the relationship between the dependent and independent variables is assumed to be linear. It is also assumed that the residuals' distribution is normal and their variance is constant across different levels of the independent variables in a linear model \cite{AssumptionsLM}. The first assumption of linearity is important; therefore, many researchers \cite{bartlett1993linear, kiernan2013natural} use Pearson correlation \cite{pearson} to select the predictor variables to model a dependent variable. Pearson's correlation, also known as the Product-Moment Correlation Coefficient (PMCC) \cite{PMCC}, is a dimensionless covariance measure. PMCC is a common statistical tool used to measure the strength and direction of the linear relationship between two variables. It can be applied if the variables are numeric, have a linear relationship, and are normally distributed. The variable that shows a higher PMCC with a response variable is assumed to result in a more efficient linear model if used as a predictor variable. Correlation and linear modeling are different but not mutually exclusive \cite{asuero2006correlation}. PMCC is mathematically defined to have a value range between -1 to 1 and its meaning is hard to comprehend \cite{brillinger2001does}. A PMCC value of -1 means a perfect negative relationship between the predictor and response variables. A +1 PMCC means that as the predictor variable increases, the response variable increases constantly. A Pearson's correlation of a value of 0 indicates that no particular pattern can be identified between the predictor and response variables. A linear model can benefit from such information because it builds a line that passes over the data points using particular criteria. If two variables have a +1 or -1 PMCC, they are expected to produce a perfect linear model with no errors when this model is evaluated. A linear model can be evaluated using evaluation metrics such as Mean Absolute Percentage Error (MAPE), Mean Absolute Error (MAE), and Root Mean Squared Error (RMSE). The lower the value of a metric, the better the model is.\\
This article aims to test if PMCC can be reliable in selecting the predictor variables for linear models. The article demonstrates the findings using built-in datasets from R namely mtcars~\cite{mtcars} and iris~\cite{iris}. The mtcars and iris datasets are two of the most widely used pre-loaded datasets in the R programming language, serving as valuable resources for statistical analysis and machine learning applications. The mtcars dataset, derived from the 1974 Motor Trend US magazine, provides information on fuel consumption and various design and performance characteristics of 32 automobiles. It includes variables such as miles per gallon, number of cylinders, horsepower, and transmission type, enabling researchers to explore relationships between car features and fuel efficiency through regression analysis. On the other hand, the iris dataset contains measurements of sepal length, sepal width, petal length, and petal width for 150 iris flowers across three different species. This dataset is commonly used to demonstrate classification techniques, as the species of each flower can be predicted based on morphological characteristics. Both the mtcars and iris datasets have become staples in the R community, facilitating the development and testing of statistical models across a variety of domains.\\
This paper is outlined as follows: First, the methodology section describes the datasets and some statistics, presents the PMCC and linear model formulas, and the error metrics used to evaluate the linear models. The results section exhibits the PMCC and then the linear model results for each dataset. Finally, a conclusion and recommendation are presented.

\section{Methodology}
Linear models are easily built models and efficient enough for various datasets. Additionally, linear models take less space than other machine learning models. A linear model can be built to represent a dependent variable Y using an independent variable X. Equation \ref{EQ1} represents the line of the linear model built using X and Y. As the equation shows, a positive perfect relationship means a constant increase in the Y value as the X value increases. A negative perfect relationship means a constant decrease in the Y value as the X value increases. A linear model consists of four components. The response (dependent) variable is denoted as Y which is the outcome that a researcher is interested in estimating its value. The predictor (independent) variable/s is denoted as X which is the variable used to predict the value of Y. $\beta$ is the slope of X when modeling Y. The intercept $\alpha$ is the value of Y when X is zero.\\
\begin{equation}
 Y = \alpha + \beta X
 \label{EQ1}
\end{equation}

\begin{figure}
  \centering
  \includegraphics[width=\linewidth]{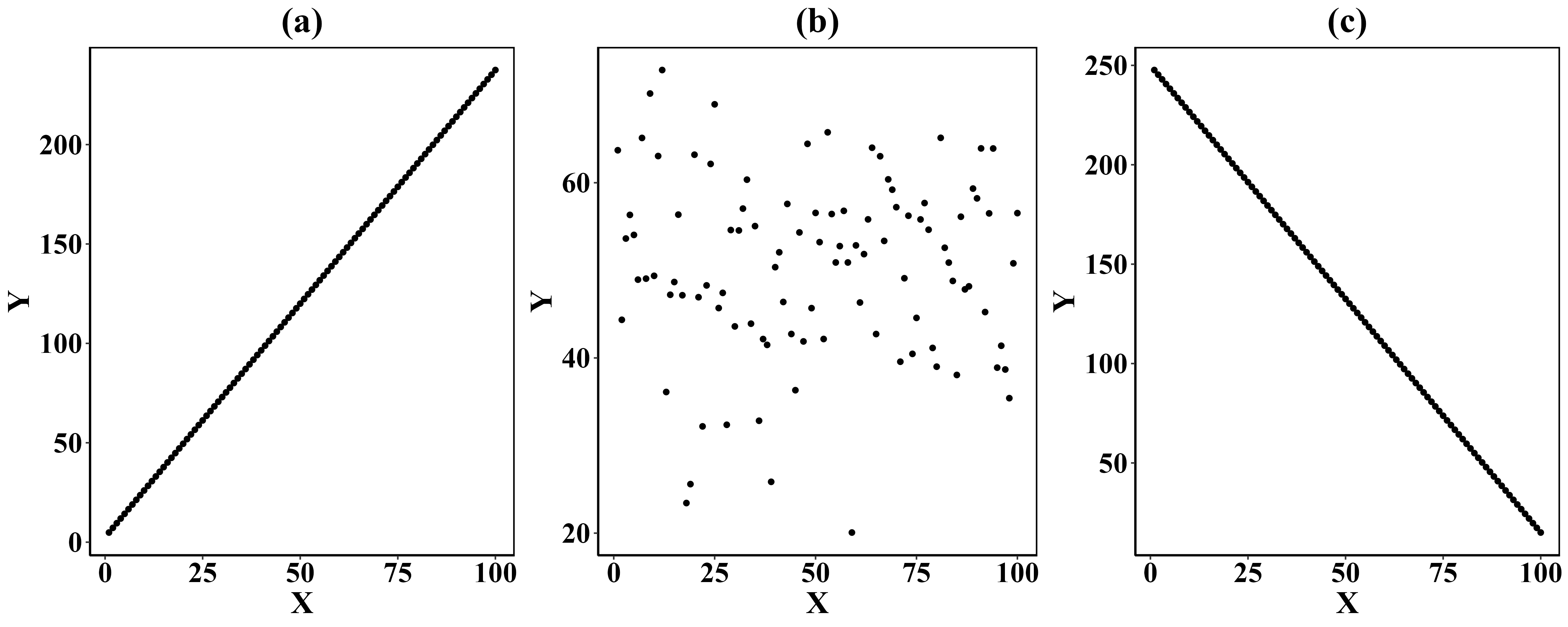}\\
  \caption{Demonstrating various types of correlations: (a) Perfect positive correlation. (b) No correlation or pattern. (c) Perfect negative correlation.}\label{Dem}
\end{figure}

Before building a linear model, it can be useful to apply a correlation measure that can assess the level of association between X and Y. Product-moment correlation coefficient (r) can be a good indicator for the level of association between the predictor and response variables. Equation (\ref{EQ2}) shows the math to calculate r for x and y. 

\begin{equation}
    r = \frac{\sum(x_i - \bar{x})(y_i - \bar{y})}{\sqrt{\sum{(x_i - \bar{x})^2} \sum{y_i - \bar{y})^2}}}
    \label{EQ2}
\end{equation}
where r is the PMCC, $x_i$ and $y_i$ are the $i^{th}$ values of each variable, and $\bar{x}$ and $\bar{y}$ are the averages of each variable.\\
Figure~\ref{Dem} shows three types of correlations, perfect positive in Figure~\ref{Dem}-a (r = 1), no correlation (r = 0) in Figure~\ref{Dem}-b, and perfect negative in Figure~\ref{Dem}-c (r = -1). PMCC can take any value between -1 and +1 depending on the association's level between the x and y.\\

\subsection{Error Metrics}
Error metrics can be used to assess the performance of linear models. MAPE, MAE, and RMSE are common metrics to assess the performance of a model when the response variable is continuous. Equations (~\ref{EQ3}) to (\ref{EQ5}) represent MAPE, MAE, and RMSE.
\begin{equation}
    MAPE = \frac{1}{n} \sum_{i=1}^{n} \left|\frac{y_i - \hat{y_i}}{y_i}\right| \times 100
    \label{EQ3}
\end{equation}

\begin{equation}
    MAE = \frac{\sum_{i=1}^{n} \left|{y_i - \hat{y_i}}\right|}{n}
    \label{EQ4}
\end{equation}

\begin{equation}
    RMSE = \sqrt{\frac{\sum_{i=1}^{n} ({y_i - \hat{y_i}})^2}{n} }
    \label{EQ5}
\end{equation}

where $\hat{y_i}$ is the predicted value corresponded to $y_i$ using $x_i$.\\

\subsection{Study Case Data}
\begin{figure}[b]
  \centering
  \includegraphics[width=\linewidth]{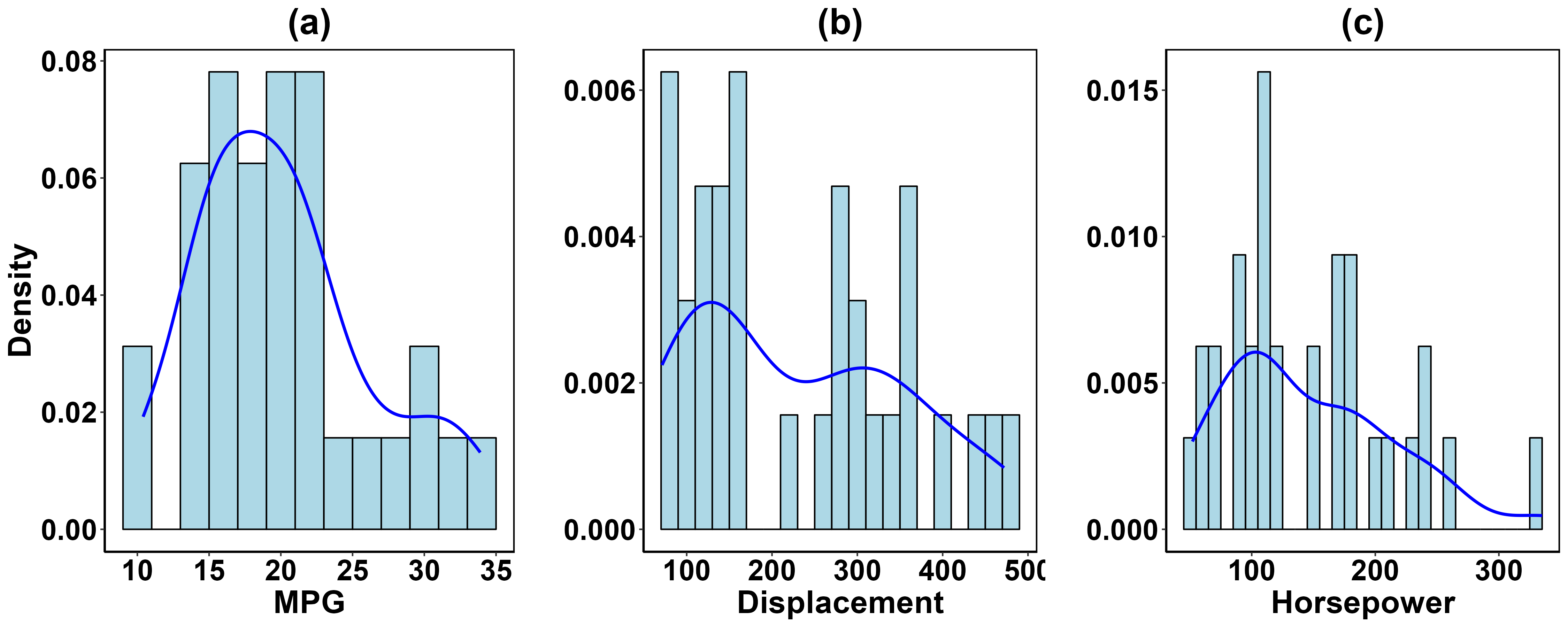}\\
  \caption{Distribution of Variables for mtcars data. a) mpg Distribution and Density Plot. b) Displacement Distribution and Density Plot. c) Horsepower Distribution and Density Plot}
  \label{hist}
\end{figure}

\begin{figure}
  \centering
  \includegraphics[width=\linewidth]{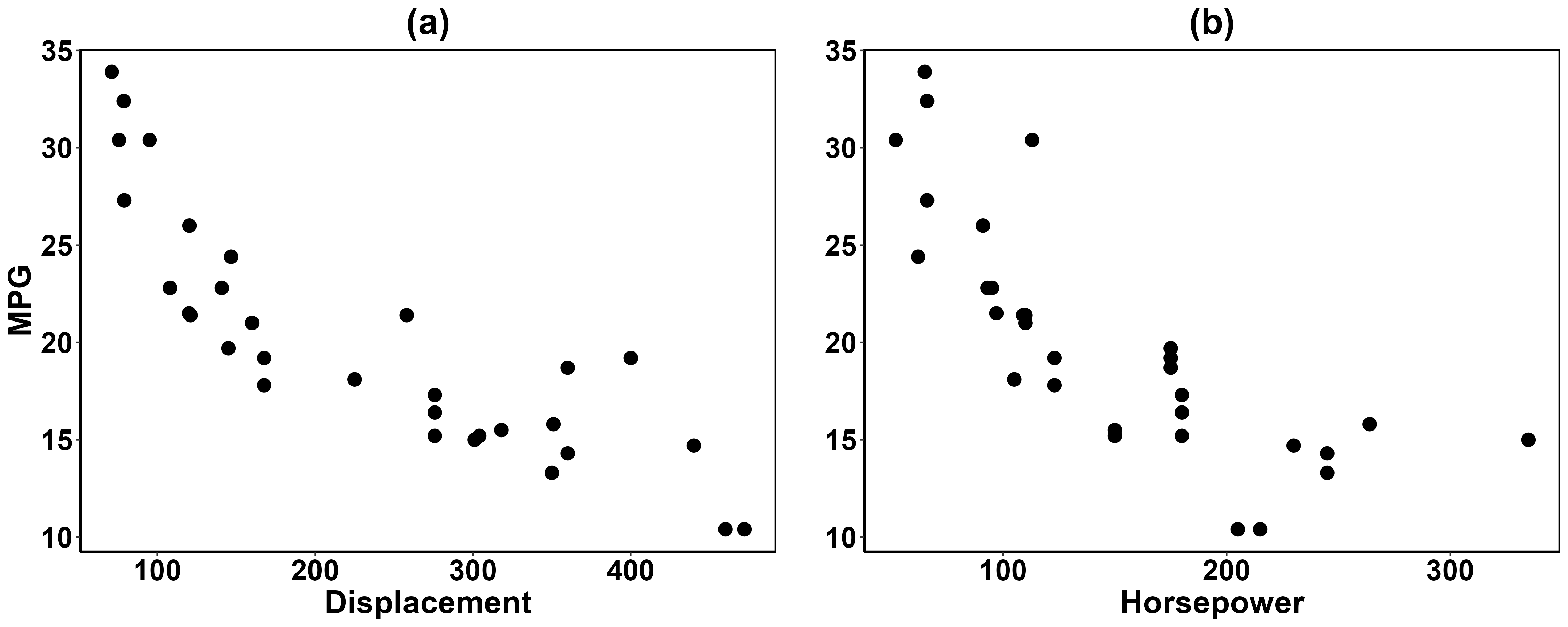}\\
  \caption{Scatter Plots for mtcars. a) The Relationship Between Displacement and MPG. b) The Relationship Between Horsepower and MPG.}
  \label{Cor1}
\end{figure}

This article demonstrates its findings using the mtcars and iris built-in datasets in R. The variables of interest in the mtcars data are displacement, horsepower, and miles per gallon. Figure~\ref{hist} shows the distribution and the density plot of the three variables. The figures reveal that the distribution of MPG and horsepower is close to normal; however, the distribution of displacement is less normal than the other two. Table~\ref{table1} shows the ranges of the MPG, displacement, and horsepower. Figure~\ref{Cor1} shows the scatter plot for the relationship between MPG and the two candidate predictor variables. The relationship is fairly linear in both cases. However, one might argue that the relationship between displacement and mpg is more linear, or at least less curvy.\\

\begin{table}
 \caption{Summary Statistics}
  \centering
  \begin{tabular}{llll}
    \toprule\\
    Variable     & Min     & Avg    & Max \\
    \midrule
    mpg & 10.40  & 20.09  & 33.90  \\
    disp     & 71.1 & 230.7   &  472.0\\
    hp     & 52.0       & 146.7  & 335.0\\
    \bottomrule
  \end{tabular}
  \label{table1}
\end{table}

\begin{figure}
  \centering
  \includegraphics[width=\linewidth]{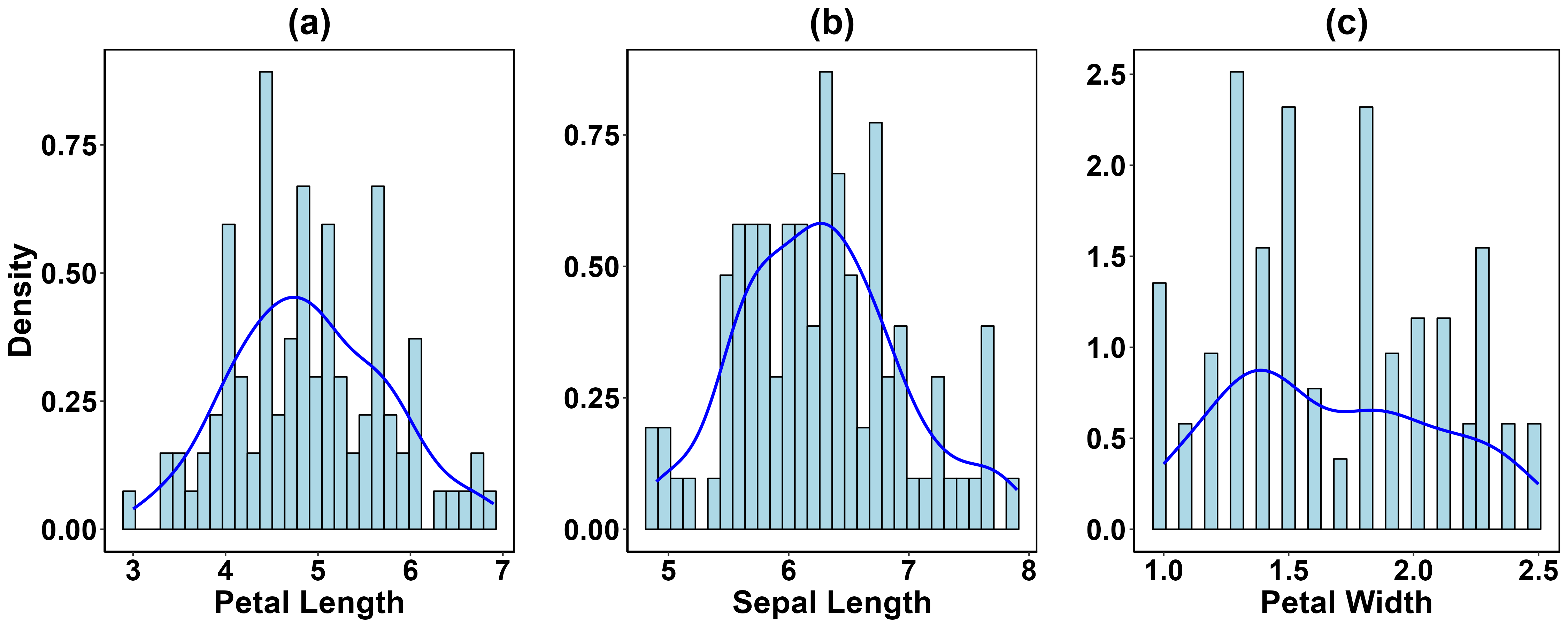}\\
  \caption{Scatter Plots for iris. a) The Relationship Between Sepal Length and Petal Length. b) The Relationship Between Petal Width and Petal Length.}
  \label{Cor2}
\end{figure}

\begin{figure}
  \centering
  \includegraphics[width=\linewidth]{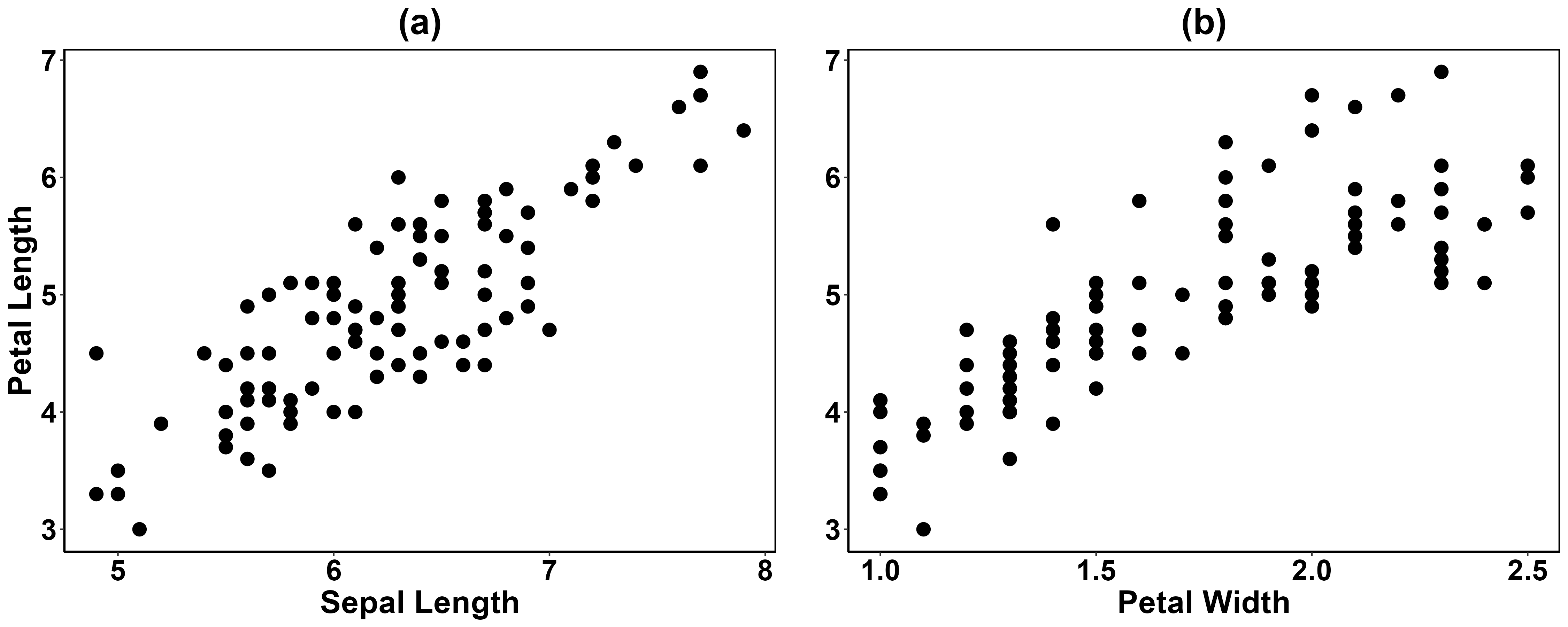}\\
  \caption{Distribution of Variables for iris data. a) petal length Distribution and Density Plot. b) sepal length Distribution and Density Plot. c) petal width Distribution and Density Plot}
  \label{Hist2}
\end{figure}

Figure~\ref{Hist2} exhibits the distribution of the petal length, sepal length, and petal width for the iris data. The first two variables show fairly normal distribution while the petal length shows two intervals, 1-1.75 cm, and over 1.75 cm. The first interval of petal width appears to be fairly normal, on the opposite of the second interval. Figure~\ref{Cor2} illustrates the relationship between the petal length and the two candidate predictor variables. Both plots (a) and (b) in Figure~\ref{Cor2} show a fairly linear relationship with some variation.

\section{Results}
This section demonstrates the findings as follows: First, PMCC is reported for a) the association between predictor and the response variables. Second, a linear model is built for each association and the evaluation metrics are reported. Finally, the results from PMCC and the linear model for each association are compared.

\subsection{mtcars Results}
The results for the mtcars data are demonstrated in this subsection as follows. 
\subsubsection{Pearson's Correlation}
As mentioned previously, PMCC is calculated to measure the correlation between mpg and displacement, and between mpg and horsepower. PMCC is found to be -0.85 and -0.78 in the two scenarios, respectively. These results demonstrate that the number of miles traveled is more associated with the displacement than the horsepower. A researcher might assume that to model the mpg, it is better to use displacement as a predictor variable. 

\subsubsection{Linear Models}
\begin{figure}
  \centering
  \includegraphics[width=\linewidth]{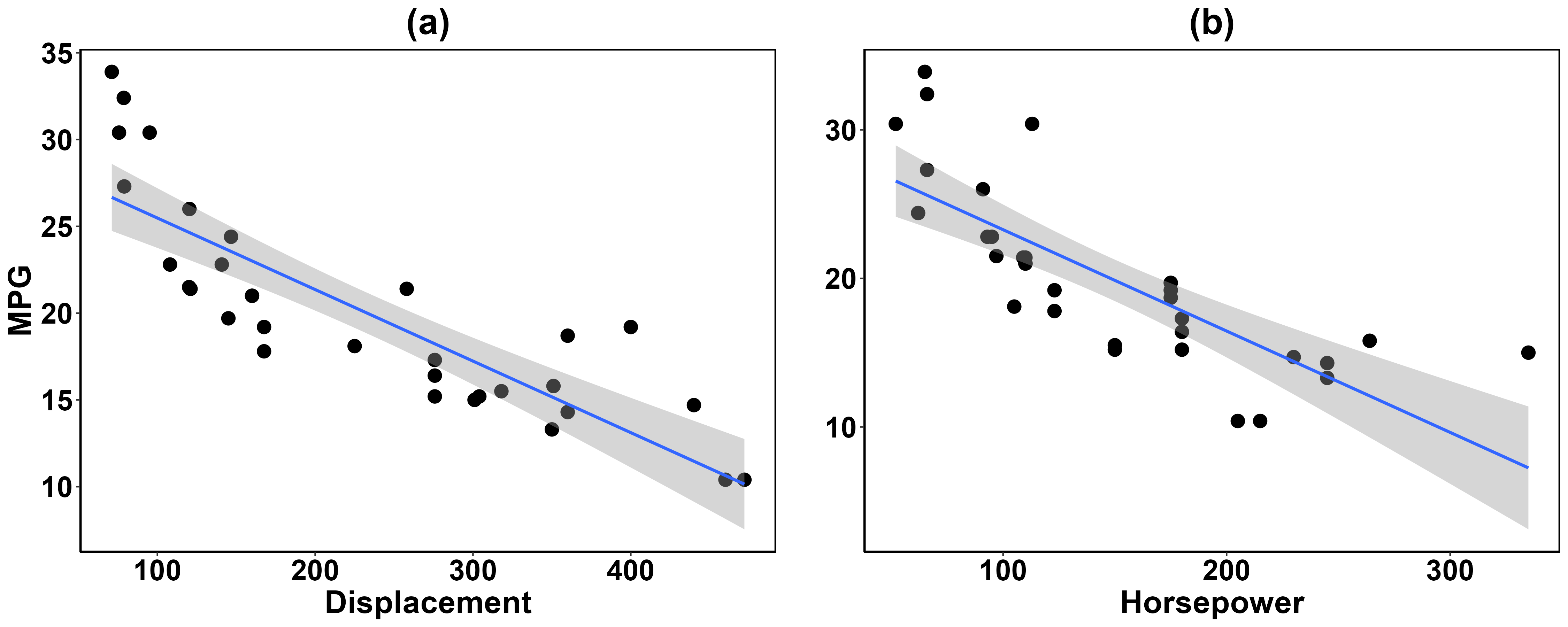}\\
  \caption{Linear models (a) for mpg as explained by displacement and its r, and (b) for mpg as explained by horsepower and its r.}
  \label{Res1}
\end{figure}

Equation~\ref{EQ6} shows the linear model used to predict the mpg using displacement. $\alpha$ and $\beta$ are called the coefficient factors and are calculated as shown in Equations~\ref{EQ7} and~\ref{EQ8}. It is worth noting the similarity between the value of $\beta$ and PMCC (i.e., r) in Equation~\ref{Cor2} and Equation~\ref{EQ7}. This can lead to a close value of $\beta$ to r. On the other hand, the $\alpha$ value is straightforward and not as significant as $\beta$. Similarly, the second model for mpg as explained by hr is built as shown in Equation~\ref{EQ9}
\begin{equation}
 mpg = \alpha + \beta * disp
 \label{EQ6}
\end{equation}

\begin{equation}
    \beta = \frac{\sum_{i=1}^{n}(x_i - \bar{x})(y_i - \bar{y})}{\sum_{i=1}^{n}{(x_i - \bar{x})^2} }
    \label{EQ7}
\end{equation}

\begin{equation}
 \alpha = \bar{y} + \beta * \bar{x}
 \label{EQ8}
\end{equation}

\begin{equation}
 mpg = \alpha + \beta * hp
 \label{EQ9}
\end{equation}

where $\alpha$ is the intercept of the linear model line between the dependent and independent variables and $\beta$ is the slope of that line. \\
These two models are shown in Figure~\ref{Res1} with the corresponding r values. The figures also show the standard error for each linear model. It is demonstrated by the Figure that a higher standard error is shown by horsepower than displacement towards mpg.

The next step is to evaluate the performance of these linear models. MAPE, MAE, and RMSE are reported in Table~\ref{table2} for each model. The first model produces a higher error than the second model, although the first model is built using displacement which is more associated with mpg than hp is.

\begin{table}
 \caption{Evaluation Metrics}
  \centering
  \begin{tabular}{llll}
    \toprule\\
    Predictor Variable    & MAPE (\%)    & MAE (Mile)   & RMSE (Mile) \\
    \midrule
    Displacement & 19  & 3.68  & 4.35  \\
    Horsepower     & 14 & 2.75   &  4.07\\
    \bottomrule
  \end{tabular}
  \label{table2}
\end{table}

The results from Table~\ref{table2} suggest that the error rate is lower when Horsepower is used to predict the mpg. In other words, when horsepower is applied, the predicted mpg value is closer to reality than the predicted values of mpg when using displacement as a predictor variable.

\subsection{iris Results}
The results for the iris data are demonstrated in this subsection as follows. 
\subsubsection{Pearson's Correlation}
For the relationship between the petal length and the sepal length, the PMCC is reported to be 0.83, whereas the PMCC between the petal length and petal width is 0.82. The two values are pretty close with a slightly higher PMCC between the petal length and sepal length. Therefore, sepal length might be assumed to be a better predictor variable for the lengths of the petals.

\subsubsection{Linear Models}
Equation~\ref{EQ10} shows the linear model used to predict the petal length using the sepal length. Similarly, the second model for the petal length as explained by the petal width is built as shown in Equation~\ref{EQ11}
\begin{equation}
 Petal Length = \alpha + \beta * Sepal Length
 \label{EQ10}
\end{equation}

\begin{equation}
 Petal Length = \alpha + \beta * Petal Width
 \label{EQ11}
\end{equation}

\begin{figure}
  \centering
  \includegraphics[width=\linewidth]{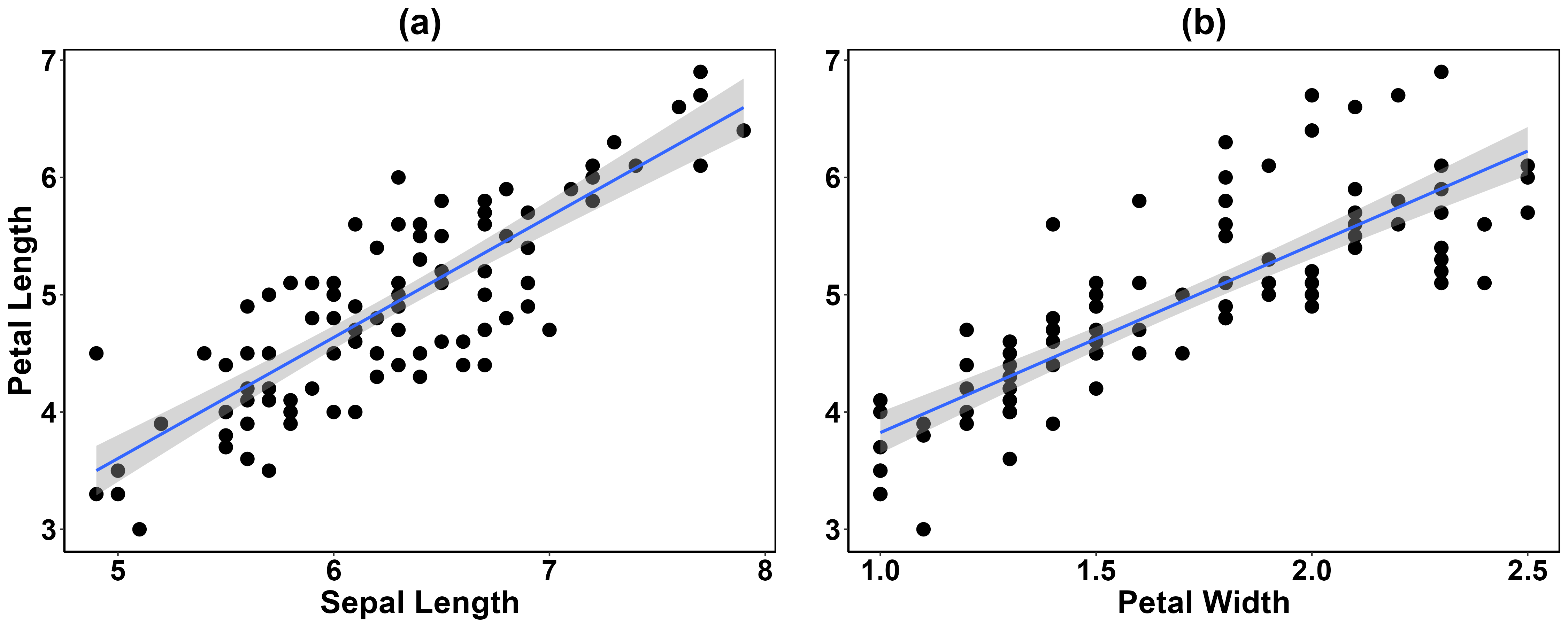}\\
  \caption{Linear models. (a) Petal Length as Explained by Sepal Length and its r, and (b) Petal Length as Explained by the Petal Width and its r.}
  \label{Res2}
\end{figure}

These two models are shown in Figure~\ref{Res2} with the corresponding r values. The figures also show the standard error for each linear model. It is demonstrated by the Figure that a slightly higher standard error is shown by the sepal length than the petal width towards the petal length.

The next step is to evaluate the performance of these linear models. MAPE, MAE, and RMSE are reported in Table~\ref{table3} for each model. For the MAPE and MAE, the first model produces a higher error than the second model, although the first model is built using sepal length which is more associated with petal length than petal width is.

\begin{table}
 \caption{Evaluation Metrics for iris Models}
  \centering
  \begin{tabular}{llll}
    \toprule\\
    Predictor Variable    & MAPE (\%)    & MAE (cm)   & RMSE (cm) \\
    \midrule
    Sepal Length & 0.073  & 0.35  & 0.43  \\
    Petal Width     & 0.070 & 0.342   &  0.460\\
    \bottomrule
  \end{tabular}
  \label{table3}
\end{table}

The results from Table~\ref{table3} suggest that the error rate is lower when petal width is used to predict the petal length. In other words, when petal width is applied, the predicted values petal length value is closer to reality than the predicted values of petal length when using sepal length as a predictor variable.

\section{Conclusion}
This article investigated the efficiency of Pearson's Product-Moment Correlation Coefficient (PMCC) to select a predictor variable to predict a response variable in linear models. PMCC is utilized to quantify the strength and direction of the linear relationship between two continuous variables. It effectively measures how closely the variables are linearly associated, providing insights into their correlation. A linear model is employed to represent the relationship between two continuous variables, indicating that these variables can predictably influence each other. This interchangeability suggests that changes in one variable are associated with changes in the other, allowing for a clear analysis of their linear correlation. Given the nature of both PMCC and linear model, it can be assumed that PMCC can be performed to select the suitable predictor variable for a linear model. This article used two built-in datasets from R to test this assumption. The mtcars and iris are common classical datasets used to demonstrate statistical facts. This research applied PMCC and linear modeling to three continuous variables in each dataset. The results indicated that PMCC does not represent the linear model relationship. Hence, it is not correct to rely on the Pearson's correlation when selecting a predictor variable for a linear model. Since each dataset is comprised of various types of variables with various distributions suggests a data-driven approach to measure the level of correlation between any two continuous variables, regardless of their relationship.

\bibliographystyle{unsrt}  
\bibliography{references}

\end{document}